\documentclass[usenatbib]{mn2e}
\usepackage{graphicx}

\newcommand{\Porb}{{P_{\rm orb}}}

\title[Synthetic Direct Impact Light Curves]{Synthetic Direct Impact Light Curves of the Ultracompact AM CVn Binary Systems V407 Vul and HM Cnc}
\author[M. A. Wood]{M. A. Wood$^{1}$\thanks{E-mail:
wood@fit.edu}\\
$^{1}$Department of Physics and Space Sciences,  Florida Institute of Technology, 150 W. University Blvd., Melbourne, FL 32901, USA\\
Department of Astrophysics/IMAPP, Radboud University Nijmegen, P.O. Box 9010,
6500 GL Nijmegen, The Netherlands}
\begin{document}

\date{Accepted 19 Jan 2009. Received  19 Nov 2008}

\pagerange{\pageref{firstpage}--\pageref{lastpage}} \pubyear{2009}

\maketitle

\label{firstpage}

\begin{abstract}
The interacting binary white dwarf (AM CVn) systems HM Cnc and V407 have orbital periods of 5.4 min and 9.5 min, respectively.  The two systems are characterized by an ``on/off'' behaviour in the X-ray light curve, and optical light curves that are nearly sinusoidal and which lead the X-ray light curves in phase by about 0.2 in both systems.  Of the models that have been proposed to explain the observations, the one that seems to require the least fine tuning is the {\it direct impact} model of Marsh \& Steeghs (2002).  In this model, the white dwarf primary is large enough relative to the semi-major axis that 
the accretion stream impacts the surface of the primary white dwarf directly without forming an accretion disc.  Marsh \& Steeghs (2002) proposed that in this situation there could be a flow set up around the equator with a decreasing surface temperature the further one measured from the impact point.  In this study, we estimate the light curves that might result from such a temperature distribution, and find them to be reasonable approximations to the observations. One unexpected result is that two distinct X-ray spots must exist to match the shape of the X-ray light curves.
\end{abstract}


\section{Introduction}
 
The ultracompact binary stars V407 Vul \citep[RX J1914.4+2456,][]{motch95,motch96} and HM Cnc \citep[RX J0806.3+1527,][]{israel99,br01} have been the subject of extensive study since their discovery.  Believed to be members of the interacting binary white dwarf (AM CVn) class, each system displays only a single photometric period, presumed in both cases to be the orbital period.  For HM Cnc, $\Porb = 321.5 \rm\ s = 5.36$ min \citep{israel99,stroh03}, and for V407 Vul, $\Porb = 569.4\rm\ s = 9.49\ min$ \citep{motch96,cropper98,stroh04}.  In a previous publication \citep[][hereafter Paper I]{dws08} we simulated the hydrodynamics of the accretion stream using the method of smoothed particle hydrodynamics (SPH). We found that the impact spot was small enough and thus the accretion stream ram pressure high enough for the specific kinetic energy of the flow to be advected and thermalized below the photosphere of the primary, such that it can be emitted ``downstream'' of the impact point in the equatorial flow that should develop.  In Paper I we discussed the physical nature of that accretion-driven flow and suggested the  
general features of any plausible photospheric temperature distribution that should result.

The observational constraints for the systems HM Cnc and V407 Vul were discussed in detail in Paper I \citep[also see][]{cropper04}.  Briefly, the systems are characterized by ``on/off'' X-ray light curves that trail the optical light curves by roughly 0.2 in phase \citep[and references therein]{bar07}, no hard X-rays are detected, the observed periods are decreasing with time at a rate consistent with gravitational radiation angular momentum loss \citep{stroh03,stroh04,rch06}, and flickering is minimal or absent \citep{bar07}.  

There are several models that have been proposed to explain the observations of these two systems: intermediate polar, polar, unipolar inductor, and direct impact.  First, in the intermediate polar model  originally suggested for V407 Vul, the observed periods reflect a white dwarf spin period \citep{motch96,israel99,nhw04}.  However, the lack of observed flickering, emission features, or any other observed (e.g., orbital) period, as well as the ``on/off'' character of the X-ray light curve, argue against this model.  Citing these issues \citet{cropper98} proposed that V407 Vul could be a helium analog of a polar (AM Her) system \citep[see also][]{ramsey02}, but the lack of observed He emission lines, polarization, and hard X-ray photons make it unlikely this model is correct for these systems \citep[e.g.,][]{israel02}.  

The unipolar inductor (UI) model \citep{wu02,dall07} is the only model without a Roche-lobe filling secondary.  Similar to the model proposed by \citet{gold69} for the Io-Jupiter system in which the conducting moon Io generates an electromotive force (EMF) as a result of the relative motion of the magnetic field of Jupiter.  The weak plasma environment allows currents to flow, driven by this EMF, and results in observable bright spots where these currents are dissipated in the atmosphere of Jupiter.  The UI model was intriguing given initial reports by \citet{ramsey00} and \citet{israel03,israel04} that the systems V407 Vul and HM Cnc (respectively) displayed X-ray and optical light curves that were nearly anti-phased. However, \citet{bar07} were able to show that in fact timing errors had crept into the previous optical/X-ray analyses, and that for both systems the optical phase leads the X-ray phase by roughly 0.2.  A roughly-constant phase offset suggests a common physical cause for both systems, but the corrected result presents a severe problem for the UI model. In the UI model it is expected that the X-ray-bright footprints of the magnetic field lines that carry the currents should be nearly on the meridian that passes through the line of centres between the two stars\footnote{Some minimal bending of the field lines will occur -- see for example Figure 3 of \citet{gold69}.}.  The expected UI geometry thus yields optical light curves anti-phased from the X-ray, assuming that the optical modulation comes from the heated face of the secondary star.  The \citet{bar07} result that the optical leads the X-ray in phase by $\sim$0.2 for both systems implies that the X-ray spots are rotated more than 90$^\circ$ away from the line of centres {\it for both systems}.  First, it appears on the face of it that this cannot plausibly be a stable minimum energy plasma current configuration in the co-rotating frame for both systems, but even if so, this geometry places the secondary star below the horizon of the X-ray spots thus removing the presumed heating source of the face of the secondary star.  \citet{bar05} discuss these and further difficulties with the model, and conclude it does not apply to these systems.

The final proposed model is the direct impact model. \citet{nelemans01} showed that if initial Roche-lobe contact and the onset of mass transfer in the AM CVn systems occurs at a short enough orbital period, the accretion stream can impact the surface of the primary white dwarf directly, and thus not form a disc.  \citet{ms02} applied this to suggest that the observations for V407 Vul could be explained by just such an Algol-like model.  Marsh \& Steeghs (2002) proposed that the bulk of the accreted matter would thermalize below the photosphere, mix, and boil to the surface at a temperature cool enough to emit soft X-rays. The temperature excess should broaden in latitude as additional atmospheric material is entrained in the equatorial flow that would be established, and the peak emission energy should shift toward the optical with distance from the upwelling point. 

Because no hard X-rays are detected from these systems, for this model to be viable the accretion stream must have sufficient ram pressure to advect the specific kinetic energy of the accreted material below the photosphere before thermalizing.  In Paper I we simulated the system geometry \citet{ms02} proposed for V407 Vul and found using SPH simulations that this constraint is satisfied.  The density in the accretion stream is roughly Gaussian about the midpoint of the flow, with a 1-$\sigma$ half width of 23 km in physical units at the impact point.  The contours of mass flux for the impact region on the surface of the primary white dwarf are well approximated by a normal bivariate (2-D Gaussian) distribution.  Fitting the isoflux contours, we found $\sigma_\phi = 164$ km (in longitude along the equator) and $\sigma_\theta = 23$ km (in latitude).  The 3-$\sigma$ ellipse encloses 99\% of the mass flux, and has an area of $A=\pi(3\cdot23)(3\cdot164)\rm\ km^2 = 1.1\times10^5\ km^2$ or a fraction of the stellar surface of $f_{3\sigma}=8.5\times10^{-5}$.  The mean mass flux within this area is $\dot M / A \approx 600\rm\ g\ cm^{-2}\ s^{-1}$ for $\dot M = 10^{-8}\ M_\odot\rm\  yr^{-1}$, the mean density of the stream is $\rho\sim2\times10^{-6}\rm\ g\ cm^{-3}$, and the velocity of the accreted mass is roughly $v\approx\sqrt{2GM_1/R_1}\approx0.01c$. 

\section{The Model}

Within the direct impact model, the location of the accretion stream impact spot is fixed in the binary co-rotating frame; the spot is on the accreting star's equator, and has an {\it impact angle} with respect to the line of centres that is a function of the mass ratio of the system and the radius (mass) of the primary white dwarf.  For most physically-plausible system parameters yielding dynamically-stable mass transfer, the impact point has an impact angle roughly in the range 80 to 120$^\circ$ \citep{bar07}.  In this geometry, the altitude of the accretion stream as viewed from the impact point is $\la45^\circ$, the bulk of the specific angular momentum is horizontal to the local stellar surface, and we expect that an equatorial flow will rapidly be established.  

The accreted material will penetrate far below the photosphere to a depth where the envelope pressure is of order the ram pressure.  For the mass transfer rates typical of AM CVn systems of such short orbital period \citep[$10^{-8}\la \dot M\la10^{-7} M_\odot\rm\ yr^{-1}$;][]{bild06}, we expect  $P_{\rm ram}\sim10^{13}\rm\ dynes\ cm^{-2}$, a pressure found at a depth of $\sim60{\rm\ km}\sim 0.006 R_1$ in a typical $0.6 M_\odot$ white dwarf of $T_{\rm eff}\approx 20,000$~K \citep{wood95}.  In a non-interacting white dwarf of this mass and temperature, the temperature, density, and degeneracy parameter at this depth are $T\sim 600,000$ K, $\rho \sim 0.1\rm\ g\ cm^{-3}$, and $\eta\approx -4$, respectively.  The fractional mass of the layer exterior to this depth is $M_{\rm shell}\sim 10^{-9} M_\star$, or roughly the mass accreted in a few days to a month at the $\dot M$ rates quoted above.  If we assume instead that the stream penetrates to a depth of where the envelope pressure is an order of magnitude higher than the ram pressure, $P_{\rm env}\approx 10^{14}\rm\ dynes\ cm^{-2}$, we find the depth to be $\sim100$ km or equivalently $10^{-8}\ M_\star$ in mass.  The temperature, density, and degeneracy parameter at this depth in the unperturbed model are $T\sim1.2\times10^6$ K, $\rho\sim 0.6\rm\ g\ cm^{-3}$, and $\eta\approx -3$, respectively.  The observed $\sim$65 eV X-ray source temperature \citep{israel03,stroh08} is thus consistent with upwelling of material from these depths.

The thermalized temperature of the accreted gas ($\sim100$ keV) will be orders of magnitude higher -- and the density orders of magnitude lower -- than the surrounding environment, leading to energetic convective overturn.  In the original \citet{ms02} model, the authors proposed that the relative phasing of the optical and X-ray light curves might be explained by a temperature profile around the equatorial band characterized by a high temperature at the upwelling point and then a decreasing photospheric temperature as the angle from the impact point increases.  Because the equatorial flow is in the same direction as the orbital motion, a distant observer will see the characteristic temperature {\it increasing} with time, culminating with the passing of the X-ray-bright spot across the face of the star -- this is what causes the optical light curves to lead the X-ray light curves in phase.
 
In the direct impact model the highest photospheric temperature should be some distance downstream of the accretion stream impact point where the upwelling occurs, and the range in latitude over which the photospheric temperatures exceed the nominal mean stellar photospheric temperatures should be of order but larger than the width of the accretion stream at impact.  Using the results obtained in Paper I, the 3-$\sigma$ profile enclosing 99\% of the mass flux has a full width of $\sim$138 km, or equivalantly $0.78^\circ$ on a white dwarf of radius $0.0144\ R_\odot$ ($M=0.5\ M_\odot$).  The length of the 3-$\sigma$ profile is 980 km, or equivalantly $5.6^\circ$. We expect that the further downstream from the impact point, the broader the zone should be in which the temperature is above the nominal mean photosphere, and the smaller that temperature excess should be.   

\begin{figure}
\begin{center}
\includegraphics[width=240pt]{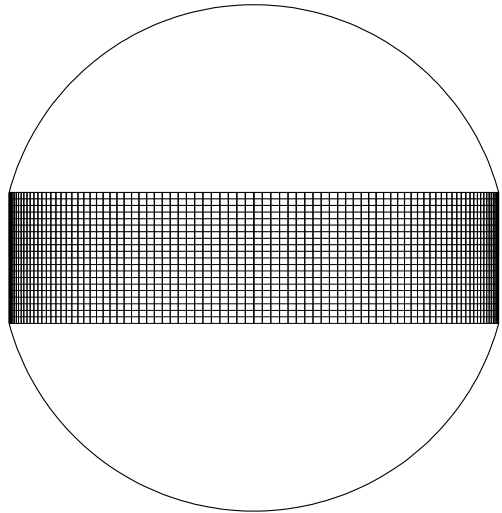}
\caption{A schematic view of the computational grid. The numerical integration is over the heavy grid represented.  The grid with lighter lines is symmetric with respect to the equatorial plane, and does not need to be computed explicitly.}
\label{fig:grid}
\end{center}
\end{figure}

We wrote a code in IDL that models a temperature distribution on a cylindrical wall that represents the equatorial band of the accreting white dwarf.  We take the latitude limits of this cylinder to be $\pm15^\circ$ for the calculations here.  A schematic representation of the computational layout is given in Figure~\ref{fig:grid}.  The actual computational grid has 360 elements in longitude ($0^\circ\le\phi<360^\circ$), and 100 elements in latitude ($0^\circ\le\theta\le5^\circ$).  Because of the symmetry with respect to $\theta=0^\circ$, we do not need to explicitly calculate the contributions for $\theta<0^\circ$ since in our model $T(\phi,-\theta) = T(\phi,\theta)$.  Although integration over a spherical surface would yield slightly different results for very low inclination angles, it would not change the overall results that we present in this preliminary study.  

In our IDL code, we integrate numerically over the visible grid elements ($180\times 100$) at each phase.  We compute for each element the blackbody spectral energy distribution for that element's temperature, correct for projection angle (including inclination) and limb darkening, and then sum over all visible elements.  Finally, the contribution from the rest of the stellar surface is included, and the integrated spectral energy distribution is convolved with the Sloan $u'g'r'i'$ filter transmission functions for direct comparison with the observational results compiled by \citet{bar07}.  For the X-ray band, we simply assume a window with 100\% transmission from 20 to 60\AA\ -- roughly the sensitivity band of the {\it Chandra} LETG spectrograph used by \citet{stroh08} in his study of HM Cnc.

Knowing that neither system is eclipsing, but otherwise uncertain about the system inclinations, we assume $i=45^\circ$ in the model.  At this inclination the cylindrical approximation for the equatorial band is still valid.  As the inclination decreases from $90^\circ$, the fraction of the visible surface made up by the equatorial band declines and the temperature contrast or latitude limits of the band must increase to yield the observed light curve amplitudes.  At very low inclinations the visibility drops for the hemisphere below the equator, and the lower latitudes of the equatorial band may be hidden (see Figures 2 and 3), again requiring an increased temperature contrast or increased latitude limits to recover the observed optical light curve amplitudes.

There are countless models for the equatorial-band temperature distribution that could be employed.  
For example, we initially implemented a model in which there was a simple linear temperature decline with distance from a compact high-temperature (X-ray bright) spot, but quickly discarded that model as it produced synthetic light curves with far too much X-ray luminosity over too great a range in longitude/phase.  Our next model included a single compact high-temperature (X-ray emitting) spot. Downstream from this spot the temperature distribution is a decreasing function of both distance from the high-temperature spot, and from the equatorial symmetry plane.  In this model the temperature falls off exponentially with longitude, and then falls off in latitude approximately as a Gaussian function that has a width that increases with longitude.  We use the standard kernel function of smoothed particle hydrodynamics to approximate the Gaussian, and let our temperature distribution within the latitude limits of the equatorial band be given by the following expression:

\begin{equation}
T(\phi,\theta) = 
      T_{\rm base} + (T_1-T_{\rm base}) e^{-\phi/\tau_\phi} W[\theta,h_\theta(\phi)]
\end{equation}
\noindent
where $T_{\rm base}$ is the assumed background temperature of the equatorial band of the primary star, and $T_1$ is the peak temperature of the distribution ($T_1\sim65$ eV).  The temperature of the remainder of the star is $T_\star$. The function $W$ is the SPH kernel function \citep{ml85}, which approximates a Gaussian distribution but is identically zero beyond twice the smoothing length $h$:

\begin{equation}
W(\theta,h) = \cases{1 - {3\over2}\left({\theta\over h}\right)^2 
									+ {3\over4}\left({\theta\over h}\right)^3 
									   & $0 \leq {\theta \over h} < 1$,\cr
									   \ \cr
						{1\over 4}\left[{2-\left({\theta\over h}\right)}\right]^3, 
								 &$1 \leq {\theta \over h} < 2$,\cr
								 \ \cr
						 0 & ${\theta \over h} \geq 2$.\cr
						 }
\end{equation}

Physically we expect that the thermalized energy from the accretion stream should spread out in latitude as a function of distance from the upwelling point, and we assume the functional form 

\begin{equation}
h_\theta(\phi) = \cases{h_0 & $\phi \le \phi_0$ \cr
									    \ \cr
								  h_0 + (h_1-h_0) \phi/\phi_1 & $\phi_0 < \phi \le \phi_1$ \cr
								     	\ \cr
									h_1 & $\phi > \phi_1$.}
\end{equation}
where in the results given below we use $h_0=1.2^\circ$ and $h_1=7.5^\circ$.  
 
Unlike our original linear-decline model, this one immediately gave results that reproduced the general observational results, including the on/off X-ray pulse shape and the optical/X-ray light curve phase offset.  However, as we discuss below, a model with a single high-temperature spot does not reproduce the width of the X-ray pulses, and therefore we were forced to introduce a second X-ray spot $\sim$65$^\circ$ upstream of the original spot, characterised by temperature $T_0$.  So in practice we typically set $\phi_0 = 65^\circ$ and  $\phi_1 = 260^\circ$, where the results are not terribly sensitive to the choice for $\phi_1$.  

\begin{figure}
\begin{center}
\includegraphics[width=240pt]{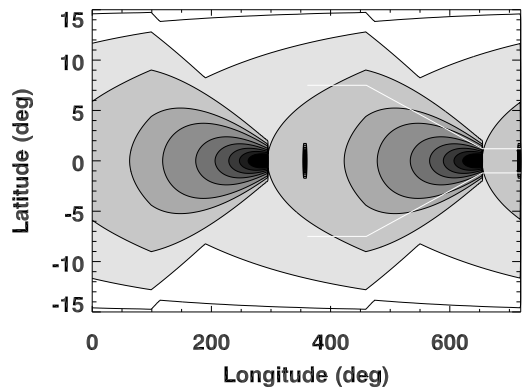}
\caption{Contour plot of two cycles of the temperature distribution representing HM Cnc.  The base temperature is $T_{\rm base}=100$ kK, and 
the contour levels are 100001 K, 100.5 kK, 110 kK, 150 kK, and 200 to 700 kK in steps of 100 kK.  The white lines plotted over the second cycle show the function $h_\theta(\phi)$.  The high-temperature feature at $\sim$$360^\circ$ and $\sim$$720^\circ$ is the impact point X-ray spot.}
\label{fig:tempconthm}
\end{center}
\end{figure}

\begin{figure}
\begin{center}
\includegraphics[width=240pt]{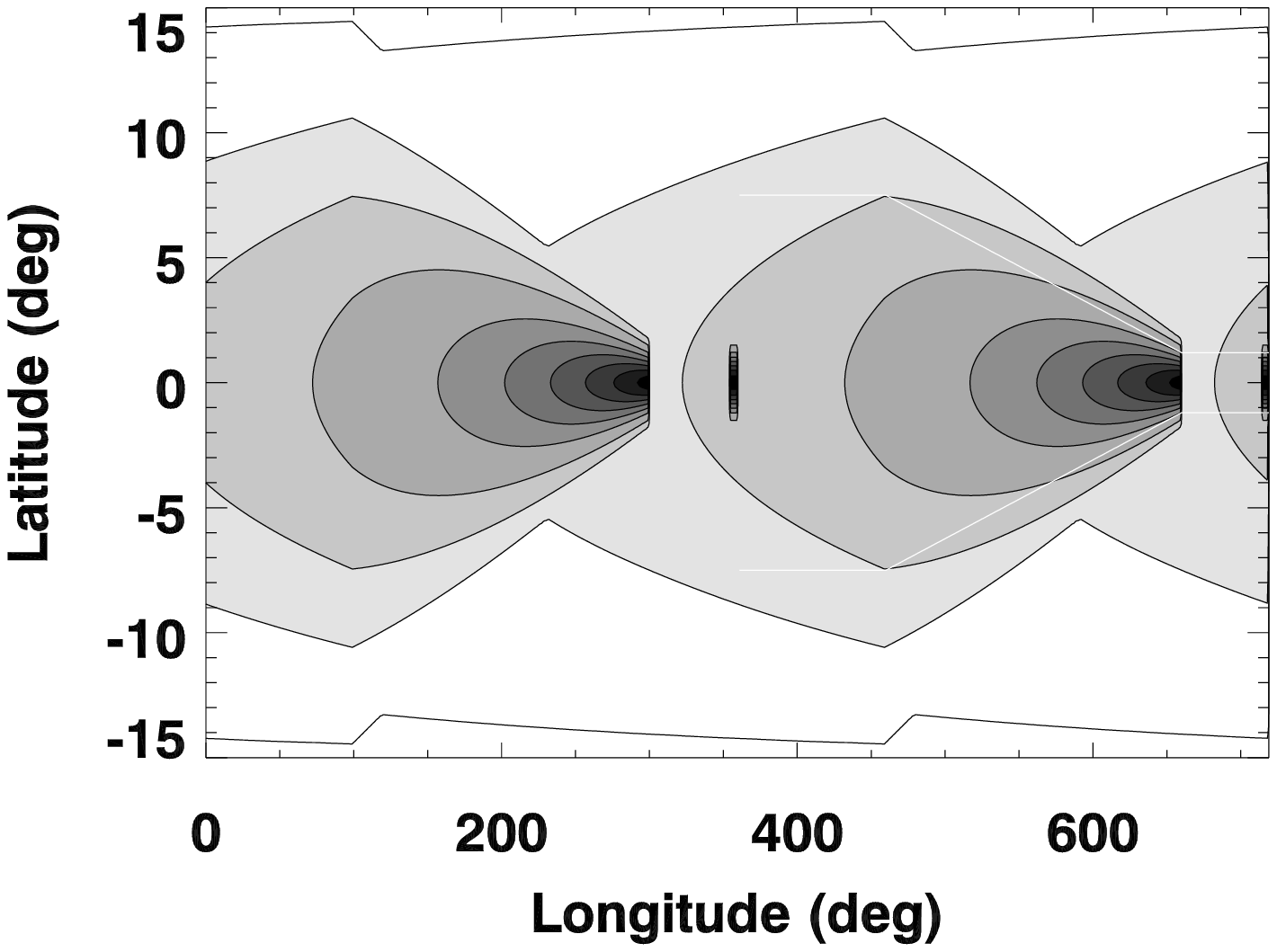}
\caption{Contour plot of two cycles of the temperature distribution representing V407 Vul.  The base temperature is $T_{\rm base}=25$ kK, and 
the contour levels are 
25010 K, 30 kK, 50 kK, and 100 to 700 kK in steps of 100 kK. The white lines plotted over the second cycle show the function $h_\theta(\phi)$.  The high-temperature feature at $\sim$$360^\circ$ and $\sim$$720^\circ$ is the impact point X-ray spot.}
\label{fig:tempcontv}
\end{center}
\end{figure}

Finally, because the e-folding length $\tau_\phi$ required to recover the observed phase offset is such that the temperature of a grid element obtained using Equation (1) extended beyond $2\pi$ exceeds that for the same element directly, we set the temperature of each grid element to be the maximum of 
$T(\phi,\theta)$ or $T(\phi+2\pi,\theta)$.  We show in Figures \ref{fig:tempconthm} and \ref{fig:tempcontv} contour plots of the temperature distributions used below to produce synthetic lightcurves representing HM Cnc and V407 Vul, respectively.

\section{The Observations}

\begin{figure*}
\begin{center}
\includegraphics[width=16.cm]{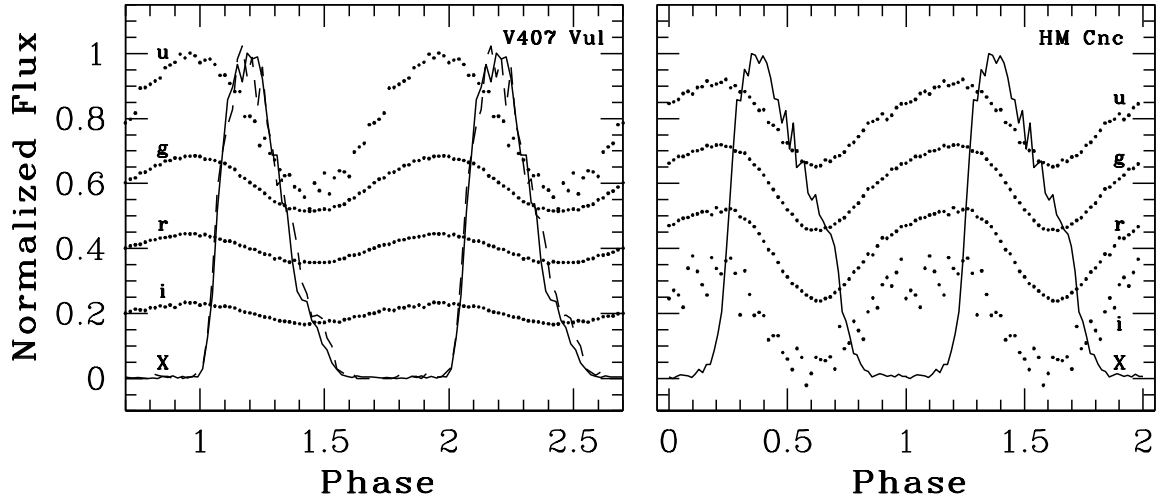}
\caption{Phase-folded light curves of V407 Vul (left-hand panel) obtained 2003 February 19 (solid line) and 2003 November 24 (dashed line) and HM Cnc (right-hand panel) as presented in \citet{bar07} and kindly sent to us by S. Barros.  Except for the HM Cnc $i'$-band data, the error bars are comparable to the size of the points, and have been omitted.  The error bars for the X-ray data are not available, but can be estimated from the scatter in the data.}
\label{fig:bar07}
\end{center}
\end{figure*}

The phase-folded light curves compiled by \citet{bar07} are shown in Figure~\ref{fig:bar07}. The authors reanalysed the {\it Chandra} X-ray data for V407 Vul obtained on 2003 February 19 (solid line) and 2003 November 24 (dashed line).  The X-ray data for HM Cnc were digitized and fit from \citet{stroh05}.  Barros et al.\ (2007) obtained the optical light curves using the ULTRACAM triple-beam CCD camera \citep{dm01} on the 4.2-m William Hershel Telescope in 2003 May and on the VLT in 2005 November.  In the course of their analysis, the authors discovered a likely error in the \citet{stroh04} V407 Vul ephemeris, and in the optical timings of \citet[][see Barros et al.\ (2007) for full details]{israel03,israel04}.

In both V407 Vul and HM Cnc, the X-ray light curves are characterised by a pulse shape that rises steeply from zero detected flux in the ``off'' state to maximum flux, and then falls more gradually from peak light.  Additionally, in both systems the phase offset between the X-ray and optical data is roughly 0.2 in phase, with the optical peak flux leading the X-ray.  Contrasting the systems, the \citet{bar07} data show that, whereas the fractional amplitudes of the V407 Vul optical light curves increase dramatically from the $r'$ to $u'$ bands, the HM Cnc fractional amplitudes are relatively constant across all optical wavelengths.  In addition, whereas the V407 Vul optical light curves are very nearly pure sinusoids, the HM Cnc light curves reveal a second-harmonic that has an amplitude 15--20\% that of the first harmonic.  The optical spectrum of V407 Vul is dominated by a late-G (or early-K) type stellar spectrum \citep{steeghs06}.  Barros et al.\ (2007) suggest this star is either a line-of-sight coincidence in a dense field, or that it is the distant member of a hierarchical triple system, but that it cannot be the direct cause of the optical or X-ray variations.  Whatever the ultimate determination, the longer-wavelength light curves are increasingly diluted from the flux from this star, causing at least in part the decreasing amplitude that is observed in longer-wavelength filters.  We note that we omit the $i'$-band light curves from the analysis below because of the increased noise in the case of HM Cnc, and because of the dilution by the G star in the case of V407 Vul.

\section{Results}

The primary goal of this project was to explore whether a plausible surface temperature distribution consistent with the direct impact model could give rise to the light curves observed for HM Cnc and V407 Vul, and in particular, whether this model could yield the phase offset and pulse shapes observed.  The results we present in this section are representative of the model output, but we caution that the solutions are not unique.

\subsection{The X-Ray Light Curves}

\begin{figure}
\begin{center}
\includegraphics[width=8.cm]{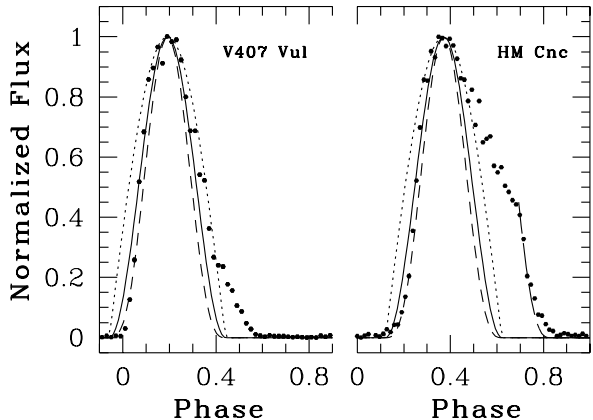}
\caption{Phase-folded X-ray light curves (filled circles) of V407 Vul (left-hand panel) and HM Cnc (right-hand panel).  Also shown are synthetic X-ray light curves calculated with three different limb-darkening laws: no limb darkening (dotted line), a linear limb darkening law with a coefficient $u_x = 1.0$ (solid line), and a quadratic limb darkening law again with $u_x = 1.0$ (short-dashed line).  Also shown is the light curve with the quadratic relation offset by 0.21 in phase (long-dashed line), demonstrating that the final decline in the X-ray light curve of HM Cnc is well matched by a second compact X-ray spot.}
\label{fig:xraysonly}
\end{center}
\end{figure}

We begin by exploring the geometry and nature of the source of the X-ray light curves in both systems. In our model, we first assumed a small X-ray-bright region presumably just downstream of the impact point, and that the temperature distribution fell off exponentially as a function of longitude, while broadening in latitude.  In Figure \ref{fig:xraysonly} we show a single cycle of the the X-ray phase-folded light curves for both systems with a simulated light curve also shown with and without limb darkening included.  For V407 Vul, we include only the 2003 February 19 data for clarity.  The simulations shown here are of a single 65 eV (750 kK) hot spot on the grid described above.  The dotted line was calculated assuming no limb darkening, the solid line was calculated assuming a linear limb darkening law $I(\mu)/I(1) = 1 - u_x (1-\mu)$, and the short-dashed line was computed with a quadratic limb-darkening law $I(\mu)/I(1) = 1 - u_x (1-\mu^2)$, where $\mu\equiv\cos(\theta)$ and where we assume $u_x = 1.0$ in both relations.  Even with this extreme limb darkening coefficient, it can be seen that for both systems the observed light curves rise more steeply than the synthetic curve.  One possible effect not included that could give the observed steep slope is a stellar surface that is non-spherical.  The hydrodynamic simulations of direct impact accretion between polytropic stars by \citet[see, e.g., their Figure 1]{motl07} show just such an effect.   

Figure \ref{fig:xraysonly} also demonstrates that the pulse width in both systems is wider than would be expected for a single X-ray-bright spot rotating into and out of view.  Interestingly, both systems show a gentler slope during part of their decline, and that slope appears to be roughly the same for both systems.  In HM Cnc, the final decline is observed to be as steep as the initial rise, which is to say as steep as a single bright spot on the stellar surface.  To demonstrate this we copied a portion of the single-spot fully-limb-darkened curve and shifted it by 0.21 in phase, and have shown this in Figure \ref{fig:xraysonly} as the long-dashed line.  We find that to match the general shape of the X-ray light curves, we were forced to include a second compact X-ray spot some 60 to 65$^\circ$ upstream of the primary X-ray spot and small enough that it does not contribute significantly to the optical flux.  We speculate that the primary spot is the upwelling point, and the secondary spot is the region around the impact spot.  While most of the specific kinetic energy of the accreted mass is advected and thermalized below the photosphere, the low-density outer sheath of the accretion stream ($\ga5$-$\sigma$ from the midline) will have $P_{\rm ram}\la P_{\rm photosphere}$ and this should yield an X-ray emitting halo around the impact point.  There will be a strong surface flow towards the impact spot in this region with speeds of order the sound speed as a result of the higher-density core of the accretion stream sweeping any and all atoms deep below the photosphere; effectively, the high-density core of the impact spot is a hole in the surface of the star into which the surrounding material will flow.  The direct impact hydrodynamic simulations of \citet{motl07}, while computed for a mass ratio of $q=0.4$ and thus higher than thought to apply to HM Cnc and V407 Vul, do indicate an upwelling point substantially displaced from the impact point.  

\subsection{HM Cnc}

\begin{figure}
\begin{center}
\includegraphics[width=240pt]{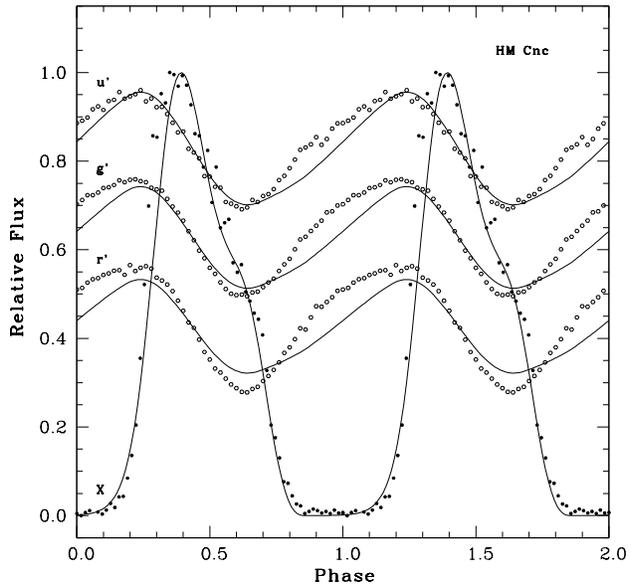}
\caption{HM Cnc data with a sample model result overplotted. The model reproduces the X-ray pulse shape, the phase offset, and the general character of the optical light curves.}
\label{fig:hmcnc}
\end{center}
\end{figure}

Figure \ref{fig:hmcnc} shows a sample model result for HM Cnc plotted over the data.  To determine suitable model parameters for each system we first find those that yield the best match ``by eye'' to the X-ray light curve, and then explore the remaining parameters to find a reasonable match to the optical light curves.
  The parameters for the HM Cnc model shown in Figure \ref{fig:hmcnc} are given in Table 1 and the contour plot of the temperature distribution is shown in Figure \ref{fig:tempconthm}, but again we emphasize that this set of parameters is almost certainly not unique and should be taken as representative, not definitive.  As an example of this, no set of parameters that we found could cleanly reproduce the excess flux in the optical between phases 0.8 and 1.0, but by artificially doubling the temperature of each grid element between longitudes 220$^\circ$ and $260^\circ$, we obtain a result which is a better approximation to the data. The physical source of this excess flux is not clear.  If the broad outlines of the current model are correct, then the source cannot be the heated face of the secondary star, as the phasing is is almost exactly wrong.  For an assumed impact angle of roughly $100^\circ$ the upwelling point is at an angle of $\sim160$ and the excess optical flux is at an angle of $\sim340^\circ$, or roughly when the back side of the secondary is towards the observer.  The hydrodynamic simulations of \citet{motl07} show some evidence for large amplitude standing waves on the surface of the accreting star after the outer layers are spun up and perhaps this could be related to the observed excess flux.  Additional work will be required to reveal the source of this radiation.

There are some physical conclusions we can make from the simulations that most closely approximate the data.  The synthetic X-ray light curve is a reasonable approximation to the data, but the data rise more steeply to maximum, perhaps because the surface of the star is slightly non-spherical.  Our experiments clearly show that two distinct X-ray spots are required to reproduce the observed X-ray pulse shape -- a single extended high-temperature ridge results in slopes of rise and fall that are shallower than those observed.  Observationally, the optical light curves all have nearly the same fractional amplitude regardless of filter, implying that the underlying source must have a very high temperature that is fluctuating, such that the optical passbands are far out on the Rayleigh-Jeans tail of the spectral energy distribution.  \citet{bild06} suggest that compressional heating should result in these stars having temperatures $T_{\rm eff} \ga 100$ kK, but we find fractional amplitudes of only $\sim$0.04 if we assume the entire stellar surface has a temperature of 100 kK, and not just the equatorial band base temperature.

\begin{table}
 \centering
  \caption{V407 Vul, HM Cnc Model Parameters.}
  \begin{tabular}{@{}llll@{}}
  \hline
   Parameter & V407 Vul & HM Cnc & Description\\  
 \hline
 $T_{\rm 0}$	& 750 kK & 880 kK & Impact spot temp \\
 $T_{\rm 1}$	& 750 kK & 750 kK & Upwelling spot temp \\
 $\phi_0$			& $60^\circ$ & $65^\circ$ & Angle between spots\\ 
 $\phi_1$			& $260^\circ$ & $260^\circ$ & Temp distribution param.\\
 $T_{\rm band}$& 25 kK & 100 kK & Boundary layer base temp \\
 $T_{\star}$	& 25 kK & 25 kK & Stellar photosphere temp\\
 $\tau_\phi$ 	& $100^\circ$ & $80^\circ$ & Temp e-folding angle\\
\hline
\end{tabular}
\end{table}

\subsection{V407 Vul}

\begin{figure}
\begin{center}
\includegraphics[width=240pt]{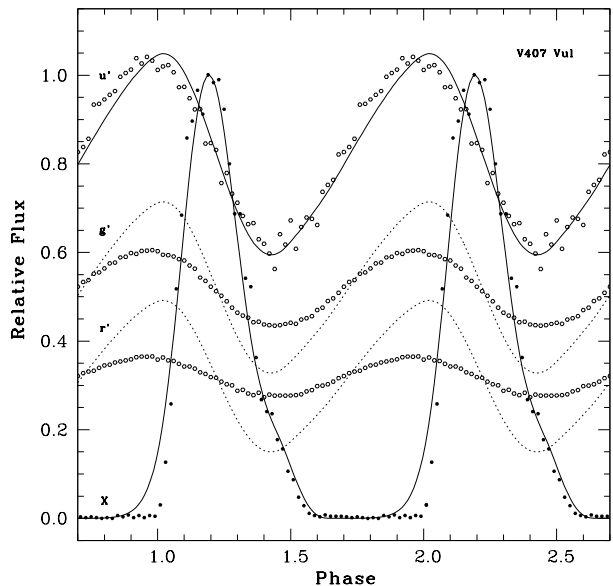}
\caption{A sample model result plotted over the observations of V407 Vul.  The model reproduces the amplitudes and shapes of the X-ray and $u'$ light curves reasonably well (solid lines). The G-star spectrum dilutes the observed amplitudes in longer passbands.}
\label{fig:v407vul}
\end{center}
\end{figure}

Figure \ref{fig:v407vul} shows a sample model result for V407 Vul, obtained using the parameters given in Table 1.  The temperature distribution is shown in Figure \ref{fig:tempcontv}.  The direct impact model provides a reasonable approximation to the observations.  As discussed above, the optical spectrum of V407 Vul is dominated by a G-type stellar spectrum, so the amplitude variations observed are not intrinsic to the direct impact scenario, and as a result are not reproduced by the simulations.  \citet{steeghs06} note that the amplitudes are consistent with a constant G-star spectrum plus a variable blue component.
For this model we show the $g'$ and $r'$ synthetic light curves as dotted lines.  In order to recover the observed phase offset for this star, we must use an e-folding length of $\tau_\phi\approx100^\circ$, but this then yields excess flux in the X-ray light curve in advance of the observed rise.  

\section{Conclusions}

We have shown that a variant of the direct impact model of \citet{ms02} can reproduce the general features observed from the ultracompact binary AM CVn systems HM Cnc and V407 Vul.  These preliminary results are promising, but should be taken as representative only. Improved fits to the observations would be possible by iterating to a minimim $\chi^2$ using differential corrections to the model parameters as is implemented in the Wilson-Devinney code \citep{wd71,wilson94}, or to allow the temperatures of the current grid points to vary freely (with or without constraints) and use a genetic algorithm or related technique \citep[e.g.,][]{goldberg89} to find solutions for the temperature distributions that match the light curves of the individual systems.  The solutions found in this way would not be unique as the number of grid points exceeds the number of data points, but additional physical insight might be gained through such a study.  This is similar to the ``firefly'' approach used by  \citet{hcr02} and \citet{bridge04} in their studies of eclipse mapping of polar accretion streams.

The primary results of this work are
\begin{enumerate}
\renewcommand{\theenumi}{(\arabic{enumi})}

\item The direct impact model of \citet{ms02} is a plausible mechanism to produce the observed X-ray and optical light curves observed in the ultracompact AM CVn binaries HM Cnc and V407 Vul.  Using a reasonable guess for the general parameters of a temperature distribution model, the optical/X-ray phase offset is reproduced, as well as the general shapes of the light curves themselves.  Subtle differences between the model results and the observations indicate that the model is still incomplete.

\item In both systems the X-ray light curves rise significantly more steeply than would be predicted for a single compact hot spot on the surface of a spherical star, even assuming an extreme limb-darkening law and coefficient for the X-ray band.  We speculate that the surface of the star may be non-spherical as a result of the ongoing accretion, and that the upwelling X-ray region in particular may have a photospheric radius slightly larger than than the average for the star.  The hydrodynamic simulations of \citet{motl07} appear to support this scenario.

\item The X-ray light curve from system HM Cnc and V407 Vul can only be reproduced by including a second X-ray-bright spot $\sim65^\circ$ upstream of the primary (brighter) X-ray region.  To fit the observed light curve, there must be two distinct spots -- a high-temperature ridge spanning this range in longitude will not yield an X-ray light curve that reproduces the observed pulse shape.  We speculate that the secondary spot is the actual impact point of the accretion stream, and that the primary spot is the region where the upwelling occurs \citep[again cf.][]{motl07}.  The X-rays from the impact region most likely originate from the low density outer regions of the accretion stream for which $P_{\rm ram}\la P_{\rm photosphere}$.  We might expect that the X-ray emission from the turbulent upwelling point might be highly variable -- considerably moreso than the X-ray emission from the impact point -- and a detailed statistical analysis of the existing X-ray time series might shed light on the viability of this proposed model.
 
\item  For the accretion rates predicted for these two systems \citep{bild06}, the ram pressure is sufficient to penetrate to a depth of order 100 km below the photosphere or $10^{-9}$ to $10^{-8} M_\star$ in mass depth -- i.e., $\log(1-M_r/M_\star)\approx8.5$, where $M_r$ is the mass contained within radius $r$.  The temperatures at these depths in the envelope of a 20 kK white dwarf model are characteristic of the 65 eV temperatures measured from X-ray spectroscopic observations.

\item The observed optical light curves of HM Cnc all have roughly the same amplitude.  To obtain this in the simulations, the base temperature of the equatorial band must be of order 100 kK, but the rest of the star must be at a lower temperature to reproduce the observed fractional amplitudes in the optical. \citet{bild06} suggest that compressional heating of the accretor should lead to mean photospheric temperatures of $T_{\rm eff} \ga 100$ kK -- marginally consistent with our simple model results.  The amplitudes of the optical-band light curves from V407 Vul increase dramatically from the $i'$ band to the $u'$ band.  Because the optical light is diluted with the G-star spectrum \citep{steeghs06} it is unknown how much of the amplitude trend in V407 Vul is the result of this dilution, and how much might be intrinsic to the accreting star's temperature distribution.  

\item In order to recover the observed phase offset between the optical and X-ray light curves in V407 Vul, the model yields non-zero X-ray flux significantly in advance of the observed rise. This suggests that the actual temperature distribution resulting from direct impact accretion may be more complex than the simple functional form we have assumed for this preliminary study.

\end{enumerate}

While the current results are intriguing, what remains to be done is to model the hydrodynamics of the direct impact scenario directly as it applies to the envelope and atmosphere of the accreting white dwarf.  Direct hydrodynamical simulations in 3D should allow the estimation of the photospheric temperature on the surface of the primary star, and from that it is straightforward to calculate an estimate of the phased light curves that should result. The hydrocode PLUTO \citep{mig07pluto} is well suited to such problems, and we are in the initial stages of implementing this project. We anticipate that explicit hydrodynamical simulations will not only reveal the true structure and dynamics of the outer layers of the accreting white dwarfs in HM Cnc and V407 Vul, but also the nature of the closely related system ES Cet \citep{ww02,esp05} where time-series spectroscopy reveals unusual features that may result from a truncated accretion disc, possibly resulting from a grazing impact of the accretion stream with the white dwarf atmosphere as the system makes the transition from a direct impact system to a normal AM CVn accretion disc.

\section*{Acknowledgments}

Thanks to Susana Barros for kindly sending an electronic copy of the observational phase-folded light curves.  
Thanks also to Gijs Nelemans for insightful comments on a draft of this manuscript, and to Danny Steeghs and Paul Groot  for useful conversations.  Thanks finally to the reviewer for comments that improved the manuscript.
The author was on sabbatical from the Florida Institute of Technology during the period of this work, and thanks Radboud University and The Netherlands Organisation for Scientific Research (NWO) for support through Visitor's Grant 040.11.046.
This research has made use of NASA's Astrophysics Data System Bibliographic Services.

\label{lastpage}


\begin{thebibliography}{}

\bibitem[\protect\citeauthoryear{Barros et al.}{2005}]{bar05} 
Barros S.~C.~C., Marsh T.~R., Groot P., Nelemans G., Ramsay G., Roelofs G., 
Steeghs D., Wilms J., 2005, MNRAS, 357, 1306 

\bibitem[Barros et al.(2007)]{bar07} Barros, S.~C.~C., et 
al.\ 2007, MNRAS, 374, 1334 

\bibitem[\protect\citeauthoryear{Bildsten et 
al.}{2006}]{bild06} Bildsten L., Townsley D.~M., Deloye C.~J., 
Nelemans G., 2006, ApJ, 640, 466 

\bibitem[\protect\citeauthoryear{Bridge et al.}{2004}]{bridge04} 
Bridge C.~M., Hakala P., Cropper M., Ramsay G., 2004, MNRAS, 351, 1423 

\bibitem[\protect\citeauthoryear{Burwitz 
\& Reinsch}{2001}]{br01} Burwitz V., Reinsch K., 2001, AIPC, 599, 522 

\bibitem[\protect\citeauthoryear{Cropper et 
al.}{1998}]{cropper98} Cropper M., Harrop-Allin M.~K., Mason 
K.~O., Mittaz J.~P.~D., Potter S.~B., Ramsay G., 1998, MNRAS, 293, L57 

\bibitem[\protect\citeauthoryear{Cropper et 
al.}{2004}]{cropper04} Cropper M., Ramsay G., Wu K., Hakala P., 
2004, ASPC, 315, 324 
	
\bibitem[\protect\citeauthoryear{Dall'Osso, Israel, 
\& Stella}{2007}]{dall07} Dall'Osso S., Israel G.~L., Stella L., 2007, A\&A, 464, 417 

\bibitem[\protect\citeauthoryear{Dhillon 
\& Marsh}{2001}]{dm01} Dhillon V., Marsh T., 2001, NewAR, 45, 91 

\bibitem[\protect\citeauthoryear{Dolence, Wood, 
\& Silver}{2008}]{dws08} Dolence J., Wood M.~A., Silver I., 2008, ApJ, 683, 375

\bibitem[\protect\citeauthoryear{Espaillat et 
al.}{2005}]{esp05} Espaillat C., Patterson J., Warner B., 
Woudt P., 2005, PASP, 117, 189

\bibitem[\protect\citeauthoryear{Gokhale, Peng, 
\& Frank}{2007}]{gpf07} Gokhale V., Peng X.~M., Frank J., 2007, ApJ, 655, 1010 
	
\bibitem[\protect\citeauthoryear{Goldberg}{1989}]{goldberg89} 
Goldberg D.~E., 1989, Genetic Algorithms in Search, Optimization and Machine Learning, Addison-Wesley, Reading

\bibitem[\protect\citeauthoryear{Goldreich 
\& Lynden-Bell}{1969}]{gold69} Goldreich P., Lynden-Bell D., 1969, ApJ, 156, 59 

\bibitem[\protect\citeauthoryear{Hakala, Cropper, 
\& Ramsay}{2002}]{hcr02} Hakala P., Cropper M., Ramsay G., 2002, MNRAS, 334, 990 

\bibitem[\protect\citeauthoryear{Israel et 
al.}{1999}]{israel99} Israel G.~L., Panzera M.~R., Campana S., Lazzati D., Covino S., Tagliaferri G., Stella L., 1999, A\&A, 349, L1 

\bibitem[\protect\citeauthoryear{Israel et 
al.}{2002}]{israel02} Israel G.~L., et al., 2002, A\&A, 386, L13 

\bibitem[\protect\citeauthoryear{Israel et al.}{2003}]{israel03} 
Israel G.~L., et al., 2003, ApJ, 598, 492 

\bibitem[\protect\citeauthoryear{Israel et al.}{2004}]{israel04} 
Israel G.~L., et al., 2004, MSAIS, 5, 148

\bibitem[\protect\citeauthoryear{Marsh 
\& Steeghs}{2002}]{ms02} Marsh T.~R., Steeghs D., 2002, MNRAS, 331, L7 

\bibitem[\protect\citeauthoryear{Marsh, Nelemans, 
\& Steeghs}{2004}]{mns04} Marsh T.~R., Nelemans G., Steeghs D., 2004, MNRAS, 350, 113 

\bibitem[\protect\citeauthoryear{Mignone et 
al.}{2007}]{mig07pluto} Mignone A., Bodo G., Massaglia S., 
Matsakos T., Tesileanu O., Zanni C., Ferrari A., 2007, ApJS, 170, 228 

\bibitem[\protect\citeauthoryear{Monaghan 
\& Lattanzio}{1985}]{ml85} Monaghan J.~J., Lattanzio J.~C., 1985, A\&A, 149, 135 


\bibitem[\protect\citeauthoryear{Motch 
\& Haberl}{1995}]{motch95} Motch C., Haberl F., 1995, ASPC, 85, 109 

\bibitem[\protect\citeauthoryear{Motch et 
al.}{1996}]{motch96} Motch C., Haberl F., Guillout P., Pakull M., Reinsch K., Krautter J., 1996, A\&A, 307, 459 

\bibitem[\protect\citeauthoryear{Motl et al.}{2007}]{motl07} 
Motl P.~M., Frank J., Tohline J.~E., D'Souza M.~C.~R., 2007, ApJ, 670, 1314

\bibitem[\protect\citeauthoryear{Nelemans et 
al.}{2001}]{nelemans01} Nelemans G., Portegies Zwart S.~F., Verbunt F., Yungelson L.~R., 2001, A\&A, 368, 939 


\bibitem[\protect\citeauthoryear{Norton, Haswell, 
\& Wynn}{2004}]{nhw04} Norton A.~J., Haswell C.~A., Wynn G.~A., 2004, A\&A, 419, 1025 

\bibitem[\protect\citeauthoryear{Postnov 
\& Yungelson}{2006}]{py06} Postnov K.~A., Yungelson L.~R., 2006, LRR, 9, 6 



\bibitem[\protect\citeauthoryear{Ramsay, Cropper, 
\& Hakala}{2006}]{rch06} Ramsay G., Cropper M., Hakala P., 2006, MNRAS, 367, L62 

\bibitem[\protect\citeauthoryear{Ramsay et al.}{2000}]{ramsey00} 
Ramsay G., Cropper M., Wu K., Mason K.~O., Hakala P., 2000, MNRAS, 311, 75 

\bibitem[\protect\citeauthoryear{Ramsay, Hakala, 
\& Cropper}{2002a}]{rhc02} Ramsay G., Hakala P., Cropper M., 2002a, MNRAS, 332, L7 

\bibitem[\protect\citeauthoryear{Ramsay et al.}{2002b}]{ramsey02} 
Ramsay G., Wu K., Cropper M., Schmidt G., Sekiguchi K., Iwamuro F., Maihara 
T., 2002b, MNRAS, 333, 575

\bibitem[\protect\citeauthoryear{Steeghs et 
al.}{2006}]{steeghs06} Steeghs D., Marsh T.~R., Barros S.~C.~C., 
Nelemans G., Groot P.~J., Roelofs G.~H.~A., Ramsay G., Cropper M., 2006, 
ApJ, 649, 382 

\bibitem[\protect\citeauthoryear{Strohmayer}{2003}]{stroh03} 
Strohmayer T.~E., 2003, ApJ, 593, L39  

\bibitem[\protect\citeauthoryear{Strohmayer}{2004}]{stroh04} 
Strohmayer T.~E., 2004, ApJ, 610, 416 

\bibitem[\protect\citeauthoryear{Strohmayer}{2005}]{stroh05} 
Strohmayer T.~E., 2005, ApJ, 627, 920 

\bibitem[\protect\citeauthoryear{Strohmayer}{2008}]{stroh08} 
Strohmayer T.~E., 2008, ApJ, 679, L109 

\bibitem[\protect\citeauthoryear{Warner, Robinson, 
\& Nather}{1971}]{wrn71} Warner B., Robinson E.~L., Nather R.~E., 1971, MNRAS, 154, 455 

\bibitem[\protect\citeauthoryear{Warner 
\& Woudt}{2002}]{ww02} Warner B., Woudt P.~A., 2002, PASP, 114, 129

\bibitem[\protect\citeauthoryear{Wilson}{1994}]{wilson94} Wilson 
R.~E., 1994, PASP, 106, 921

\bibitem[\protect\citeauthoryear{Wilson 
\& Devinney}{1971}]{wd71} Wilson R.~E., Devinney E.~J., 1971, ApJ, 166, 605 

\bibitem[\protect\citeauthoryear{Wood}{1995}]{wood95} Wood 
M.~A., 1995, NLP Vol. 443: White Dwarfs, 443, 41

\bibitem[\protect\citeauthoryear{Wu et al.}{2002}]{wu02} Wu 
K., Cropper M., Ramsay G., Sekiguchi K., 2002, MNRAS, 331, 221 


\end{thebibliography}
\end{document}